\documentclass[preprint,12pt]{elsarticle}

\usepackage{amssymb}
\usepackage{amsmath}

\usepackage{tcolorbox}
\newtcolorbox{mybox}{colback=blue!5!white, colframe=blue!75!black, boxrule=0.5mm, arc=2mm}

\usepackage{tabularx,booktabs}
\newcolumntype{Y}{>{\raggedright\arraybackslash}X}
\usepackage{hyperref}
\usepackage[nameinlink,capitalize]{cleveref}
\usepackage{subcaption}

\usepackage{lineno}
\usepackage{draftwatermark}
\SetWatermarkText{Preprint not peer-reviewed}
\SetWatermarkScale{2.7}      
\SetWatermarkColor[gray]{0.85}

\begin{document}

\begin{frontmatter}



\title{U-Net–LSTM with incremental time-stepping for robust long-horizon unsteady flow prediction}


\author{Blaise Madiega} 

\affiliation{organization={Mechanical Engineering, Université Laval},
            addressline={1065 Avenue de la Médecine}, 
            city={Quebec City},
            postcode={G1V0A6}, 
            state={Quebec},
            country={Canada}}

\author{Mathieu Olivier} 

\affiliation{organization={Mechanical Engineering, Université Laval},
            addressline={1065 Avenue de la Médecine}, 
            city={Quebec City},
            postcode={G1V0A6},
            state={Quebec},
            country={Canada}}

\begin{abstract}
Transient computational fluid dynamics (CFD) remains expensive when long horizons and multi-scale turbulence are involved. Data-driven surrogates promise relief, yet many degrade over multiple steps or drift from physical behavior. This work advances a hybrid path: an incremental time-stepping U-Net–LSTM model that forecasts unsteady dynamics by predicting field updates rather than absolute states. A U-Net encoder–decoder extracts multi-scale spatial structures, LSTM layers carry temporal dependencies, and the network is trained on per-step increments of the physical fields, aligning learning with classical time marching and reducing compounding errors. The model is designed to slot into solvers based on projection methods (such as SIMPLE, PISO etc), either as an initializer that delivers a sharper first guess for pressure–velocity coupling or as a corrective module that refines provisional fields. Across representative test cases, the approach improves long-term stability (\textbf{54.53-84.21} \% reduction of cumulative errors) and preserves engineering metrics, integral and averaged quantities, more reliably than standard learning baselines. These properties make it a plausible component of hybrid CFD–ML pipelines designed to accelerate unsteady simulations without compromising quantitative fidelity.

\end{abstract}


\end{frontmatter}



\newpage
\section{Introduction}
Numerical prediction of unsteady flows now sits at the heart of modern engineering, enabling design optimization, analysis of complex phenomena, and informed decision-making. Yet, finite-volume discretization combined with time stepping becomes expensive, as horizons lengthen or turbulence sets in \cite{Patankar1980}. Capturing multiple scales forces fine meshes and small time steps, driving wall-clock time upward \cite{Ferziger2002}. In segregated CFD solvers, the pressure correction step remains a major bottleneck to convergence \cite{Issa1986}. Machine learning (ML) provides a promising way to accelerate these computations while preserving quantitative accuracy \cite{Duraisamy2019}. Recent reviews have outlined both the practical applications and inherent limitations of such approaches \cite{Brunton2020}.

Work on spatiotemporal predictors spans several families. U-Net captures multi-scale spatial structures, which are useful for complex flow fields \cite{Ronneberger2015}; LSTM (long short term memory) models preserves long-range temporal dependencies \cite{Hochreiter1997}; and ConvLSTM folds convolution into the recurrent cell to handle image sequences coherently \cite{Shi2015}. In fluid mechanics, such pipelines forecast the evolution of a cylinder wake at a fraction of the computational cost of classical solvers \cite{Lee2019}, and on more turbulent configurations, learned predictors reproduce global dynamics over nontrivial horizons \cite{Hasegawa2020}.

A complementary approach uses graph-based models on unstructured meshes. These methods can learn physics-consistent updates for complex systems \cite{SanchezGonzalez2020} and produce predictions that generalize across different geometries \cite{Pfaff2021}.
In parallel, neural operators learn mappings between continuous fields at nearly constant computational cost. Examples include the Fourier Neural Operator for parametric PDEs \cite{Li2021FNO} and DeepONet for nonlinear operators using branch and trunk networks \cite{Lu2021DeepONet}.
Physics-constrained autoencoders capture PDE dynamics in compact latent spaces \cite{Geneva2020}. Generative models trained on parametric families can synthesize realistic flow fields at lower cost \cite{Kim2019}, while volumetric super-resolution methods enhance small-scale features without losing temporal coherence \cite{Xie2018}.

Targeted machine-learning interventions tackle the main computational bottlenecks in CFD. Convolutional networks have been used to accelerate pressure projection in Eulerian simulations \cite{Tompson2017}, while data-driven predictors embedded in segregated solvers reduce pressure-iteration overheads \cite{Sousa2024}. Integrating learned guidance directly into finite-volume schemes yields consistent performance gains on standard benchmarks \cite{Levie2024}. At the turbulence scale, end-to-end learning frameworks have shown the ability to speed up unsteady computations without degrading global flow statistics \cite{Kochkov2021}.

A persistent challenge remains: the gradual drift that accumulates in autoregressive predictions. \textit{Teacher forcing}, a common technique used in autoregressive time series forecasting, trains the model by supplying the true previous state instead of its own prediction, thereby speeding up convergence but creating a train–test mismatch that amplifies error accumulation over long horizons \cite{Bengio2015}. Sequence-level learning helps reduce this mismatch, yet maintaining long-term stability is still difficult \cite{Ranzato2016}. The \textit{DAgger} framework highlights how the choice of deployment policy affects the rate at which errors compound \cite{Ross2011}. Adding physics-based penalties can enforce consistency with governing equations and improve realism, though at the cost of more expensive training \cite{Raissi2019}. Data-driven discretizations better aligned with numerical schemes can reduce local errors, but long-horizon drift often persists \cite{BarSinai2019}. Beyond visual quality, maintaining reliable spectra and integral quantities remains essential for engineering applications \cite{Taira2017}.

Generalization adds another layer of difficulty. The Fourier Neural Operator achieves fast inference, yet often struggles to extrapolate across meshes or geometries without adaptation \cite{Li2021FNO}. DeepONet handles nonlinear operators more flexibly but remains sensitive to its chosen basis and sensor locations \cite{Lu2021DeepONet}. Super-resolution methods can reconstruct fine structures, though they rarely guarantee stability across long-term predictions \cite{Fukami2019}. As several reviews emphasize, ensuring adherence to physical constraints and energy fidelity is crucial \cite{Brunton2020}. Temporal autoencoders and recurrent models such as LSTMs have proven effective on chaotic systems, offering accurate short-term predictions but gradual degradation over longer horizons \cite{Vlachas2018}.

Beyond algorithmic stability, data availability and computational costs constitute persistent bottlenecks. While high-fidelity simulations such as DNS  and LES offer ground truth for the training of ML models, they demand prohibitive resolution and resources \cite{Moin1998}. Conversely, models trained on more accessible RANS or LES data often struggle to generalize across different flow regimes \cite{Ling2016}. Although specific techniques, like learned subgrid closures or in-situ training, can improve accuracy; they frequently introduce significant computational overhead \cite{Beck2019, MacArt2021}. Furthermore, architectural challenges remain: while Neural Operators excel on structured grids \cite{Li2021FNO}, their robustness on irregular domains is still an open question \cite{Lu2021DeepONet}. Ultimately, whether using these methods or complementary approaches like sparse model discovery \cite{Brunton2016}, reliable application demands the validation of global quantities rather than relying solely on local accuracy \cite{Taira2017}.

Altogether, key priorities emerge to fully leverage ML in CFD: limiting long-horizon error growth, preserving global flow metrics, and keeping inference costs practical \cite{Kochkov2021}. Modeling residuals instead of absolute outputs simplifies the optimization of models \cite{He2016}. In the context of unsteady flows, this naturally translates to predicting temporal increments rather than full fields, aligning with the logic of classical time-stepping schemes \cite{Patankar1980}.

To overcome challenges mentioned previously, our study introduces a U-Net–LSTM approach for transient flow prediction, paired with an incremental strategy that estimates field increments instead of absolute values to mitigate long-horizon drift \cite{He2016}. It also delineates where a spatiotemporal predictor best serves as an initial guess or a corrector within a CFD-ML hybrid solver \cite{Issa1986}.

The major contributions of our approach can be summarized as follows:
\begin{enumerate}
    \item \textbf{Mitigation of error accumulation:} By predicting increments instead of absolute values, our model ensures better temporal consistency and limits the exponential growth of errors, a significant issue in classical LSTM architectures.

    \item \textbf{Reduced dependence on massive training datasets:} Unlike traditional deep learning models that require large databases for efficient generalization, our method works effectively with smaller datasets due to its localized learning capacity.

    \item \textbf{Alignment with CFD numerical methods:} This strategy na- \newline turally follows the numerical solvers that employ time discretization schemes, thereby facilitating a smooth integration with existing CFD simulations. By predicting increments, the model can serve not only as a standalone predictor but also as a corrector in a hybrid CFD solver.

\end{enumerate}

\section{Methodology}
\subsection{Dataset Generation}
\label{subsec:dataset_generation}

\section{Methodology}
\subsection{Dataset Generation}
\label{subsec:dataset_generation}

The following sections detail the three distinct flow configurations used to generate the training data for the proposed framework. It is important to note that, as these datasets are derived from numerical simulations, they inherently contain discretization errors associated with the underlying grid and numerical schemes. However, the ultimate objective of this work is not to eliminate these errors via pure inference, but rather to utilize the trained model as a predictor (or \textit{initial guess}) coupled with a traditional CFD solver. This hybrid approach aims to accelerate the solver's convergence while maintaining physical rigor.

\subsubsection*{Test case 01: Von Karman Vortex Street: 2D Cylinder Case}
\label{subsubsec:von_karman_2d}

The dataset for the Von Karman vortex street behind a 2D cylinder was generated using CFD simulations carried out with \texttt{OpenFOAM} \cite{OpenFOAM2013}. The geometry consists of a 2D domain with a circular cylinder of diameter \( D \) placed symmetrically. Boundary conditions include symmetry planes at the top and bottom, a velocity inlet with uniform inflow velocity \( U_{\infty} \), and a pressure outlet with fixed pressure. The flow parameters are defined by a Reynolds number \( Re = 100 \) and an incompressible, Newtonian fluid. The numerical setup employs the \texttt{pimpleFoam} solver for transient, incompressible flows, with a time step $ \Delta t U\infty / D = 0.05 $  ensuring a Courant number \( Co < 1 \) for stability, and a mesh of \( 9440 \) cells to resolve the vortex shedding dynamics.

\subsubsection*{Test case 02: Kelvin–Helmholtz Shear Layer (2D)}
\label{subsubsec:kelvin_helmholtz_2d}

The Kelvin-Helmholtz (KH) shear layer is a canonical benchmark for unsteady shear flow dynamics: it exhibits roll-up, vortex pairing, and multi-scale mixing in a geometry-free, periodic setting. Consequently, the KH instability serves as an ideal test case (challenging because it is highly sensitive to numerical dissipation/dispersion, thus exposing long-horizon error accumulation) for evaluating coupled CFD-ML frameworks.

This test case follows the cylinder wake (test case 01) protocol but targets a compressible shear layer that exhibits roll-up, pairing, and mixing. The dataset is generated with an in-house finite-volume solver \cite{Mocz2020} for the compressible Euler equations on a periodic square domain \(\Omega=[0,1]^2\), discretized by a uniform mesh of \(N\times N\) cells with \(N=128\) (cell size \(\Delta x = 1/N\)). The working gas is ideal with \(\gamma=5/3\), and the initial state consists of counter-flowing streams separated by a finite-thickness strip:
All quantities below are written in non-dimensional form using  the reference length \( \sigma \) and the reference transverse velocity amplitude \( w_0 \).
\begin{equation}
(\rho^\star,u^\star,v^\star,p^\star)(x^\star,y^\star,0)=
\begin{cases}
(2,\; +0.5,\; v_0^\star(x^\star,y^\star),\; 2.5), 
& |y^\star-0.5|<0.25,\\[2pt]
(1,\; -0.5,\; v_0^\star(x^\star,y^\star),\; 2.5), 
& \text{otherwise},
\end{cases}
\end{equation}
with a transverse perturbation seeding KH growth,
\begin{equation}
\begin{cases}
v_0^\star(x^\star,y^\star)
= \dfrac{v_0(x,y)}{w_0}
= \sin(4\pi x^\star)\Big[
\exp\!\Big(-\tfrac{(y^\star-0.25)^2}{2}\Big)
+\exp\!\Big(-\tfrac{(y^\star-0.75)^2}{2}\Big)
\Big],\\[6pt]
w_0 = 0.1,\\[4pt]
x^\star = \dfrac{x}{\sigma},\qquad 
y^\star = \dfrac{y}{\sigma},\\[4pt]
\sigma = \dfrac{0.05}{\sqrt{2}}.
\end{cases}
\end{equation}

Spatial fluxes are computed using a local Lax–Friedrichs/Rusanov approximate Riemann solver with linear MUSCL-type face extrapolation (no slope limiting unless otherwise noted). Time integration is explicit with a variable step constrained by a CFL factor \(C_{\mathrm{CFL}}=0.4\),
\begin{equation}
\Delta t^\star
= C_{\mathrm{CFL}}\;
\min_{i,j}\frac{\Delta x^\star}{\,c^\star_{i,j}
+\sqrt{u_{i,j}^{\star 2}+v_{i,j}^{\star 2}}\,},
\qquad
c^\star=\sqrt{\frac{\gamma p^\star}{\rho^\star}},
\end{equation}
and both directions are periodic. The simulation advances to \(t^\star_{\mathrm{end}}=4.0\) with outputs saved every \(\Delta t^\star_{\mathrm{out}}=0.02\). Stored fields \((u,v,\rho)\) are post-processed into per-step increments \((\Delta u,\Delta v,\Delta \rho)\) using \(\Delta(\cdot)^n=(\cdot)^{n+1}-(\cdot)^n\), providing the supervision targets for the incremental time-stepping U-Net–LSTM.

\subsubsection*{Test case 03 : Fully Developed Turbulent Channel Flow (3D)}
\label{subsubsec:channel_3d}

For this use case, a Direct Numerical Simulation (DNS) was conducted using the \texttt{PyCalc solver} \cite{Davidson2020}, a finite volume method (FVM) code written in Python capable of DNS, LES, and DES (including \( k-\omega \) DES and PANS \( k-\epsilon \)). The solver uses second-order central differencing (CDS) in space and Crank-Nicolson for time integration. Synthetic anisotropic fluctuations were generated from a precursor RANS simulation using the Explicit Algebraic Reynolds Stress Model (EARSM) to compute the Reynolds stress tensor, with integral length scales \( L_{int} \) in the range \( 0.1 \cdot h/2 < L_{int} < 0.3 \cdot h/2 \), where \( h \) is the channel half-height. The flow was simulated at a friction Reynolds number \( Re_{\tau} = u_{\tau}h / \nu = 550 \), where \(u_{\tau}\) is the friction velocity and \(\nu\), the kinematic viscosity. Boundary conditions included Dirichlet conditions (no-slip walls) at the top and bottom, and periodic conditions at the inlet, outlet, front, and back boundaries to model an infinite domain in the streamwise and spanwise directions.

\subsection{Training Processes}
\label{subsec:training_processes}

\paragraph{U-Net-LSTM Architecture}

\begin{figure}[h!]
    \centering
    \includegraphics[scale=0.27]{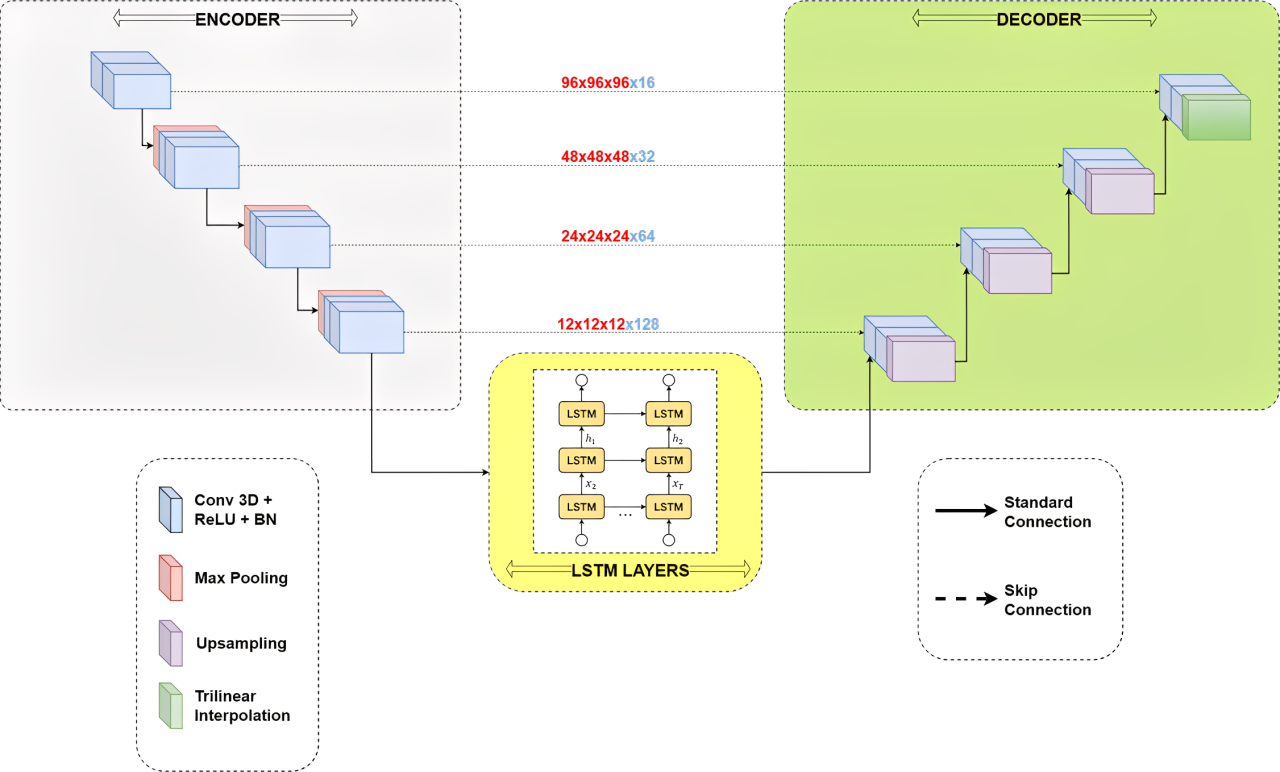}
    \caption{U\textendash Net--LSTM architecture: U\textendash Net for multiscale spatial feature extraction and LSTM for sequence modeling in unsteady flow analysis.}
    \label{fig:unet_lstm_architecture}
\end{figure}

Unsteady flow dynamics, marked by space–time variations of the velocity fields \( (u, v, w) \) and pressure \( p \), call for models that capture spatial correlations and temporal dependencies in tandem. Hybrid models that combine convolutional neural networks (CNNs) with long short–term memory (LSTM) units are well suited to this objective. In the present work, the U-Net-LSTM architecture (Figure~\ref{fig:unet_lstm_architecture}) leverages the complementary strengths of both families: the U-Net’s encoder–decoder with skip connections extracts multiscale spatial features efficiently, while stacked LSTM layers propagate long-range temporal structure across frames.

Originally developed for biomedical image segmentation, U-Net has proven effective on complex flow fields thanks to its resolution-preserving skip connections and multi-resolution context \cite{Ronneberger2015}. 

Building on this capability, integrating LSTM layers within the U-Net pipeline has been shown to improve long-horizon temporal prediction in fluid applications. In particular, a deep U-Net-LSTM framework achieved robust time-sequenced hydrodynamics prediction for the SUBOFF AFF-8 configuration, highlighting the benefits of coupling multiscale spatial encoders with sequence-aware temporal modules \cite{Hou2022UNetLSTMSUBOFF}. Together, these components provide a solid foundation for modeling turbulent unsteady flows in a manner consistent with the underlying physics.

\paragraph{Increments Prediction Approach}
\label{par:incremental_prediction}

\begin{figure}[h!]
    \centering
    \includegraphics[scale=0.25]{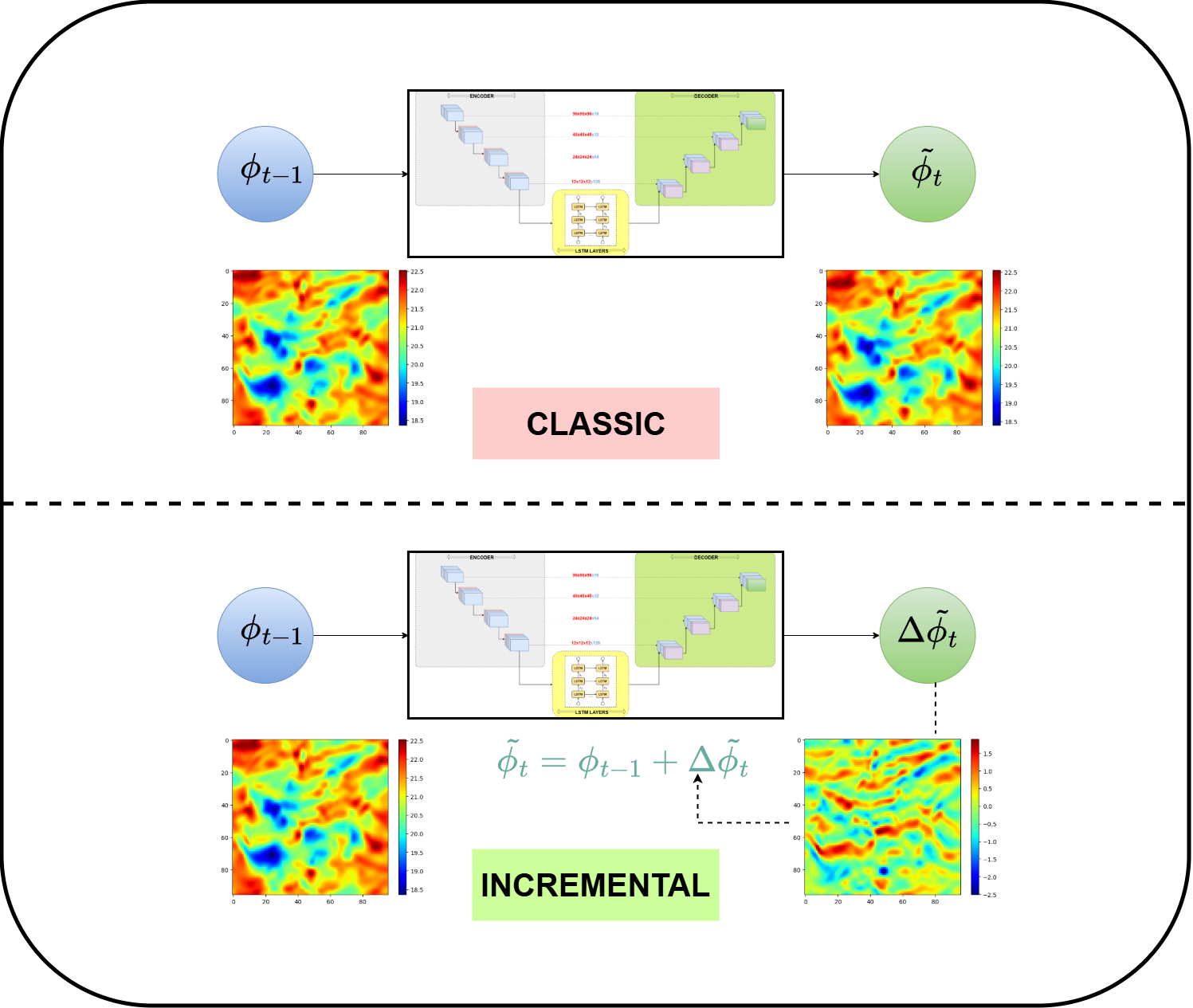}
    \caption{Schematic illustration of the two prediction approaches: (left) the \textbf{classic} approach, where the model U-Net-LSTM (illustrated at Figure~\ref{fig:unet_lstm_architecture}) predicts absolute values of physical fields; (right) the \textbf{incremental} prediction approach, where the model predicts velocity field increments at each timestep.}
    \label{fig:classic-vs-incremental}
\end{figure}

The approach developed in this work, shown in Figure~\ref{fig:classic-vs-incremental}, relies on predicting the increments of the velocity fields \( (\Delta u, \Delta v, \Delta w) \) and pressure \( (\Delta p) \) rather than their absolute values. This strategy, inspired by incremental time-stepping schemes used in CFD, reduces error accumulation over long temporal sequences. 

Formally, at each time step \( t \), the neural model predicts the increments \( \Delta \phi_t \) for each physical quantity \( \phi \) (where \( \phi \in \{u, v, w, p\} \)) from the previous state \( \phi_{t-1} \). This prediction is expressed as:
\begin{equation}
\Delta \phi_t \approx \text{NN}_{\theta}(\phi_{t-1}),
\label{eq:increment_prediction}
\end{equation}
where \( \text{NN}_{\theta} \) denotes the neural network parameterized by \( \theta \). The approximate state at time \( t \) is then obtained by integrating the predicted increment:
\begin{equation}
\phi_t \approx \phi_{t-1} + \Delta \phi_t.
\label{eq:state_update}
\end{equation}
This formulation offers major advantages:

\begin{itemize}
    \item \textbf{Physical consistency}: it aligns the model with the temporal discretization schemes used in CFD, ensuring better adherence to the underlying physics.
    \item \textbf{Error reduction}: because increments are smaller in magnitude than absolute values, they limit error propagation over long prediction horizons; as shown in the results in \Cref{sec:results}.
    \item \textbf{Integration with CFD solvers}: it eases future integration with traditional CFD pipelines, where predicted increments can act as corrective updates in a PIMPLE/PISO loop.
\end{itemize}

\paragraph{Incremental Learning}
\label{par:incremental_training}

In this study, we also investigate an incremental learning strategy to optimize the use of computational resources, particularly for three-dimensional turbulent flows, where dataset size can grow rapidly. In such settings, storing the full-time history of all steps for model training quickly becomes impractical due to memory and storage constraints.

The incremental learning approach processes data sequentially, updating the model parameters at each time step or in small mini-batches, rather than loading the entire dataset into memory. It is important to note that this training strategy, focused on data management and optimization, is distinct from the increments prediction approach detailed previously in \Cref{par:incremental_prediction}; however, both methods can be applied jointly. This method is efficient over batch training through:

\begin{itemize}
    \item \textbf{Reduced memory footprint}: it enables handling very large datasets by avoiding loading all data into memory at once.
    \item \textbf{Improved scalability}: it is well suited to long-duration simulations, where the number of time steps can reach several thousands or tens of thousands, thereby enabling efficient resource management and facilitating advanced techniques like in-situ learning or continuous adaptation.
\end{itemize}

\section{Experimentation}
\label{sec:experiments}

\Cref{tab:hp_global} consolidates the principal data, architecture, and training settings. This section details (i) the rationale behind the target selection and temporal windowing, (ii) the UNet–LSTM design and training protocol, and (iii) the hyperparameter optimization procedure carried out with Optuna \cite{Optuna} that yielded the final configurations reported in \Cref{tab:hp_global}.

\paragraph{Implementation notes}
The following definitions clarify the architecture and training hyperparameter notation:

\begin{enumerate}
    \item \textbf{Temporal window ($T$)}: $T$ denotes the number of sequential time steps processed by the model's recurrent (LSTM) layers during training. Given our single-step supervision setup, we use $T=1$ for all test cases.
    \item \textbf{Network width ($Start$-$Dim$)}: $Start$-$Dim$ specifies the base width of the UNet architecture; encoder depths follow the filter sequence $\{Start$-$Dim, 2 \times Start$-$Dim, 4 \times Start$-$Dim, \ldots\}$ filters per level (e.g., $\{16, 32, 64, \ldots\}$).
    \item \textbf{Convolutional blocks ($\texttt{num\_convs}$)}: $\texttt{num\_convs}=2$ indicates two convolutional layers per block for both 2D and 3D variants, unless hyperparameter optimization (via Optuna \cite{Optuna}) selects a different value under budget constraints.
    \item \textbf{Learning Rate ($\eta$)}: $\eta$ is the initial learning rate used for the Adam optimizer.
    \item \textbf{Decay schedule ($Step, \gamma$)}: This column specifies the learning rate decay schedule, applied using a \textbf{StepLR scheduler}. $\gamma$ is the decay ratio and $Step$ is the number of epochs after which the learning rate is multiplied by $\gamma$.
    \item \textbf{Data window ($\texttt{data\_cuttof\_start} / \texttt{data\_cuttof}$)}: These parameters specify the effective temporal window used for training, notably excluding initial transient phases (e.g., $t<1500$ for test case 01).
\end{enumerate}

\paragraph{Data windowing and targets}
A single-step supervision \(T{=}1\) is adopted to emulate CFD solver rollouts while concentrating temporal memory in the LSTM bottleneck. In 2D cases, absolute-field targets (\textit{classic}) prioritize fidelity in statistically stationary regimes; thus, test case 01 discards the initial \(t<1500\) steps to remove bias before the von Karman regime.

\begin{table}[h]
\centering
\caption{Global hyperparameters by test case}
\footnotesize
\setlength{\tabcolsep}{2pt}
\renewcommand{\arraystretch}{1.1}
\begin{tabularx}{\linewidth}{l c c c c c c c c Y}
\toprule
\textbf{Case} & \textbf{Input size} & \textbf{Channels} & $\mathbf{T}$ & \textbf{Batch} & \textbf{Epochs} & $\boldsymbol{\eta}$ & \textbf{Step,$\boldsymbol{\gamma}$} & \textbf{Start} & \textbf{Notes} \\
\midrule
Test case 01 & $128\times256$ & $[u,v,p]$      & 1 & 32 & 100 & $10^{-2}$ & 50,\;0.1 & 16 & $1500$ timesteps \\
Test case 02 & $128\times128$ & $[u,v,\rho]$   & 1 & 32 & 100 & $10^{-2}$ & 50,\;0.1  & 16 & $200$ timesteps \\
Test case 03 & $96^3$         & $[u,v,w,p]$    & 1 & 4  & 100  & $10^{-3}$ & 100,\;0.1 & 16 &  $200$ timesteps \\
\bottomrule
\end{tabularx}
\label{tab:hp_global}
\end{table}

\paragraph{Model and training protocol}
The UNet–LSTM couples multiscale spatial encoders/decoders (skip connections preserve sharp shear layers and pressure–velocity coupling) with a compact temporal state at the neck to capture slow-manifold dynamics even with \(T{=}1\). The base width (\texttt{Start-Dim}=16) doubles per level; two convolutions per block (\texttt{num\_convs}=2) enlarge the receptive field without inflating parameters. the optimizer \textsc{Adam} is selected for transient, multi-channel fields due to its robustness to heterogeneous gradient scales across variables and depths; this avoids brittle learning-rate tuning typical of pure SGD in unsteady regimes. The mean squarred error \textsc{MSE} serves as a physically interpretable \(L_2\) baseline that supports cross-case comparability. A step learning rate (\texttt{StepLR}) schedule executes a single decade drop (see step and \(\gamma\) in \Cref{tab:hp_global}), earlier in shorter 2D runs (test case 01/02), later for 3D (test case 03) to prevent premature annealing.

\paragraph{Hyperparameters' optimization with Optuna}
Final settings in \Cref{tab:hp_global} are the best trials from the Optuna study. The objective minimizes validation RMSE averaged across predicted channels (equal weight), computed on a chronologically held-out slice to avoid leakage. The search space in \Cref{tab:optuna_space} is hardware-aware and constrained by memory (the device used for experiments is an \textbf{NVIDIA RTX A4000 GPU} with \textbf{16 GB of memory} \cite{NVIDIA2025}):

\begin{table}[h]
\centering
\caption{Optuna optimization search space}
\footnotesize
\setlength{\tabcolsep}{4pt}
\renewcommand{\arraystretch}{1.05}
\begin{tabularx}{\linewidth}{@{}lX@{}}
\toprule
\textbf{Hyperparameter} & \textbf{Search space / notes} \\
\midrule
Initial LR \(\eta\) & \(\log_{10}\eta \in [-4,-2]\) (log-uniform). \\
Start-Dim & \(\{16,32\}\) (categorical); \texttt{num\_convs} \(\in \{2,3\}\) with parameter-budget penalty. \\
Batch size & \(\{4,8,16,32\}\) (filtered by GPU memory and input size). \\
\texttt{StepLR} & step \(\in \{50,100,200,400\}\), \(\gamma \in \{0.1,0.2\}\). \\
LSTM \texttt{num-layers} (Test case 03) & \(\{1,2\}\). \\
\bottomrule
\end{tabularx}
\label{tab:optuna_space}
\end{table}

Each trial trains for the full epoch budget listed in \Cref{tab:hp_global}, with pruning after a warm-up (10\% of epochs) if the median of past trials' RMSE is not outperformed. To stabilize comparisons, gradient clipping at 1.0 and a fixed seed are applied; the best configuration is then re-trained on the training split with early-stopping patience fixed to 10\% of the epoch budget and finally evaluated on the untouched test slice.

\paragraph{Data splits and metrics}
A chronological split is used to reflect deployment and \textbf{reduce model dependency on a massive dataset}: 50\% of the sequence is reserved for testing, and the remaining 50\% is used for model training (80\% training / 20\% validation). Reported metrics include (i) \textbf{field-wise root mean squared errors (RMSE)}, and (ii) \textbf{drag and lift coefficients} \((C_d, C_l)\) for the cylinder and \textbf{Reynolds-stress statistics} for the 3D channel. These complement \Cref{tab:hp_global} by quantifying long-horizon behavior that hyperparameters alone cannot guarantee.

\section{Results and discussion}
\label{sec:results}
\subsection{Speed of learning convergence and model accuracy}
\begin{figure}[htbp]
    \centering
    \begin{minipage}{0.3\textwidth}
    \centering
    \includegraphics[width=\linewidth]{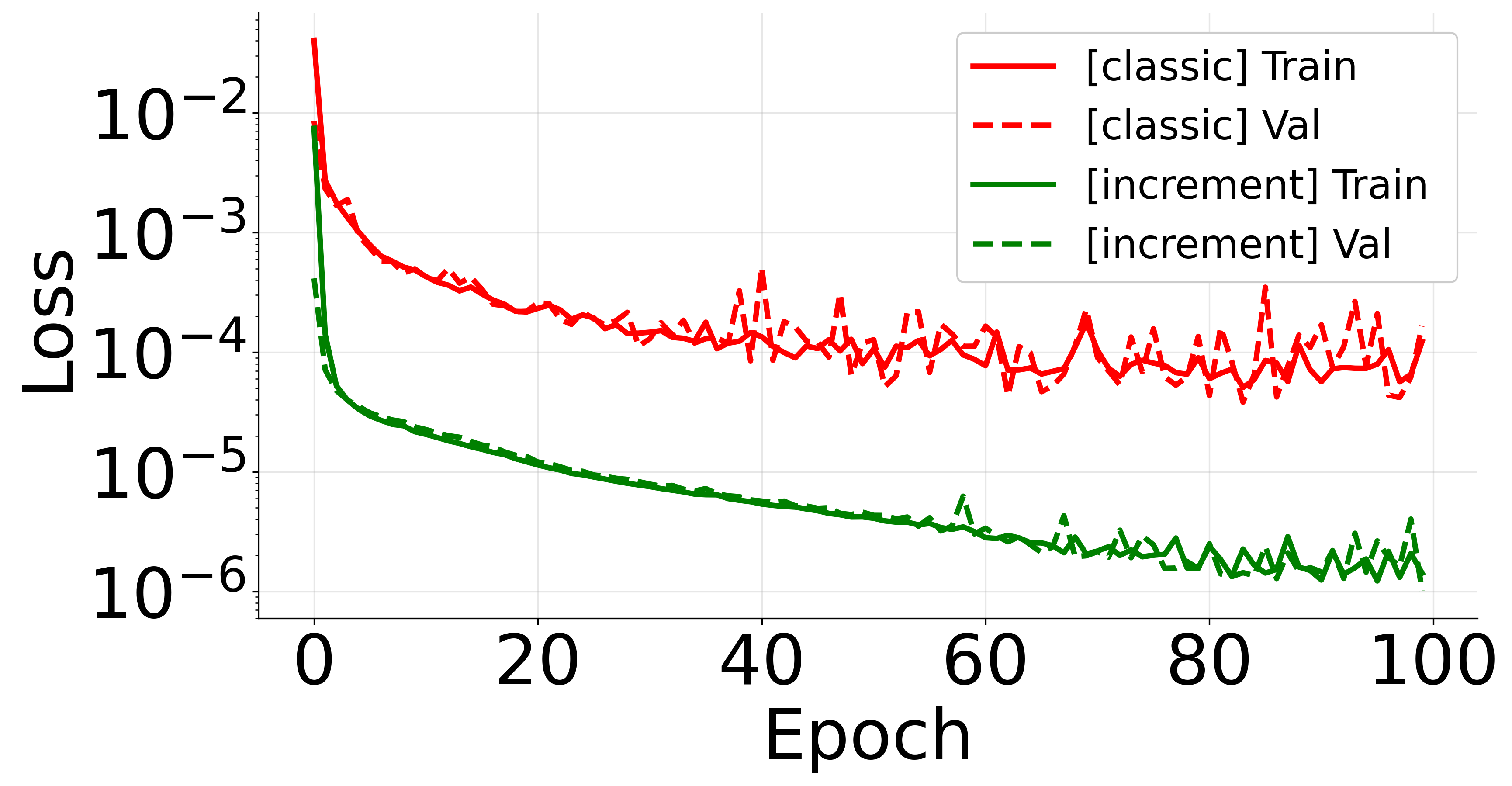}
    \subcaption{}
    \end{minipage}
     \hfill
    \begin{minipage}{0.3\textwidth}
        \centering
        \includegraphics[width=\linewidth]{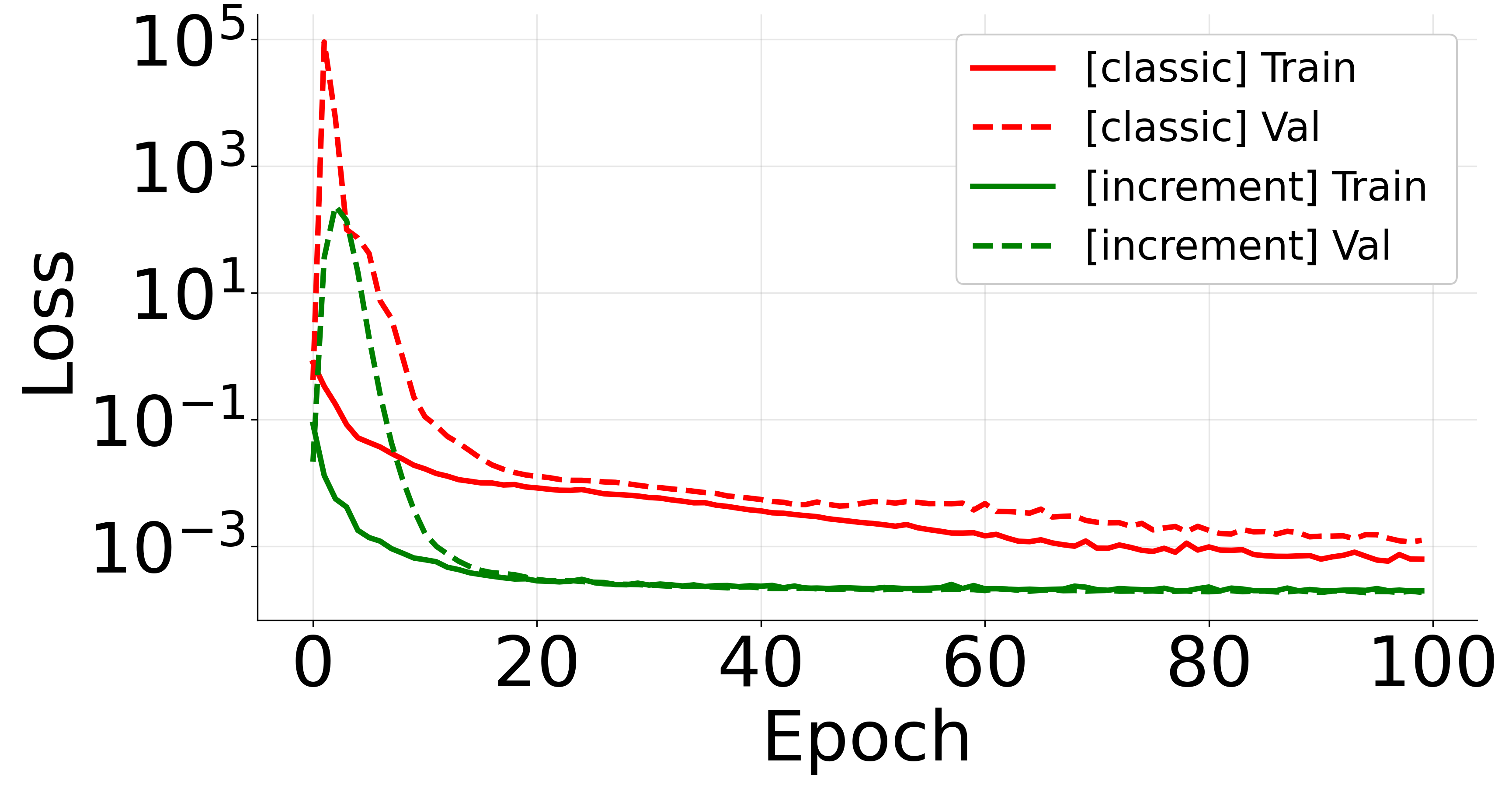}
        \subcaption{}
    \end{minipage}
    \hfill
    \begin{minipage}{0.3\textwidth}
        \centering
        \includegraphics[width=\linewidth]{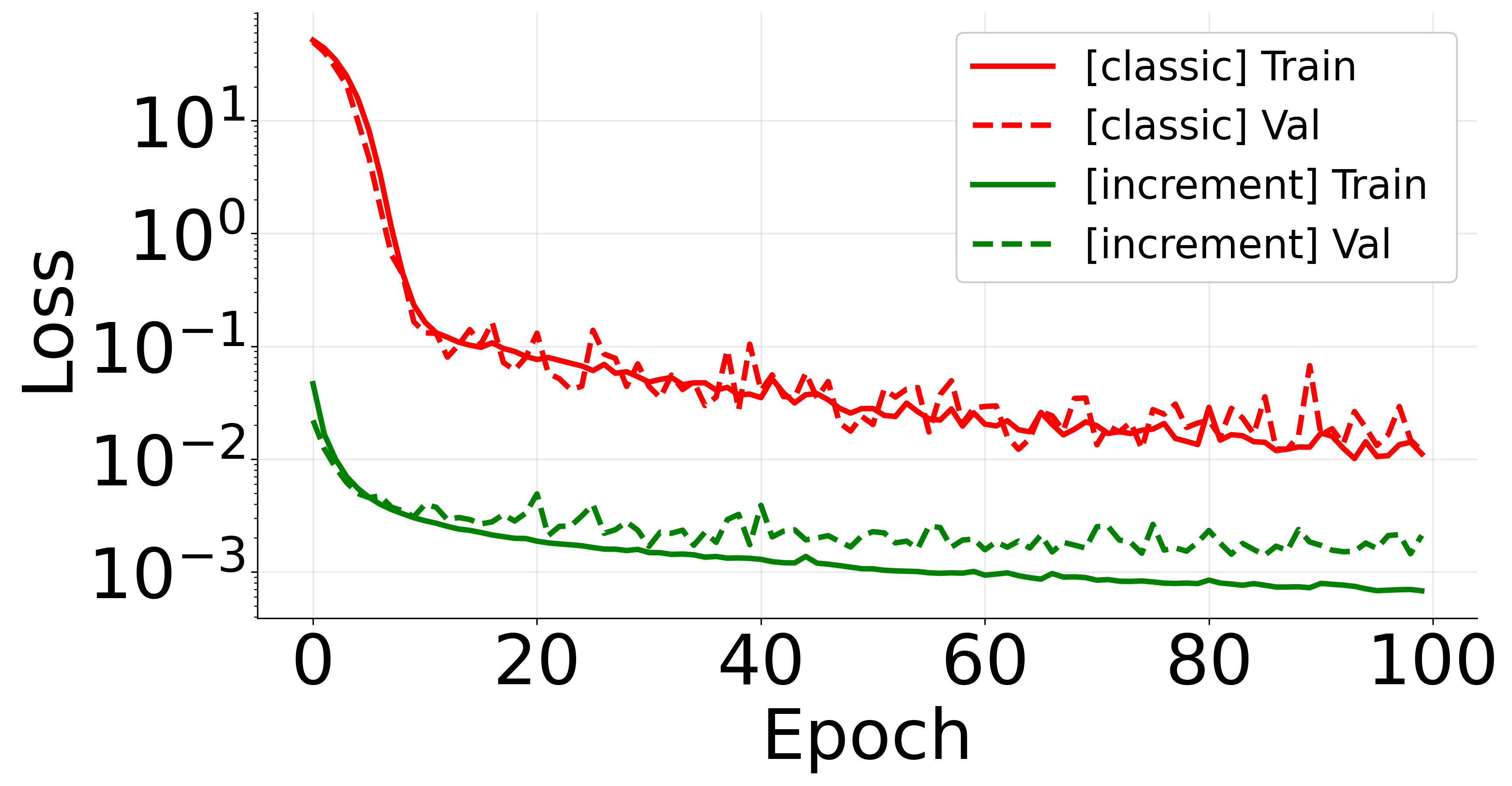}
        \subcaption{}
    \end{minipage}
    \caption{Comparison of train/validation losses over learning epochs: (a) Test case 01: Cylinder; (b) Test case 02: Kelvin–Helmholtz Shear Layer, and c)Test case 03: 3D Channel.}
    \label{fig:speed_convergence}
\end{figure}

\paragraph{Speed of learning convergence}
Across all cases, the \emph{incremental} target converges markedly faster than \emph{classic} absolute-field prediction. In the test cases 01 and 02, validation loss plateaus around 20 epochs, whereas the classic setting requires substantially more iterations to reach a comparable level; in the 3D channel configuration (test case 03) the incremental model stabilizes after 60 epochs. This rapid convergence is particularly valuable for hybrid CFD–ML solvers embedded in time-stepping workflows: it limits training overhead so that the compute cost of learning does not erode the net speed-up expected from ML-accelerated simulations. Moreover, the incremental runs attain lower terminal training/validation losses, indicating a smoother optimization landscape and more effective use of the \texttt{StepLR} decay. These trends are visible in panels a–c of \Cref{fig:speed_convergence}.

\begin{figure}[htbp]
    \centering
    \begin{minipage}{0.32\textwidth}
    \centering
    \includegraphics[width=\linewidth]{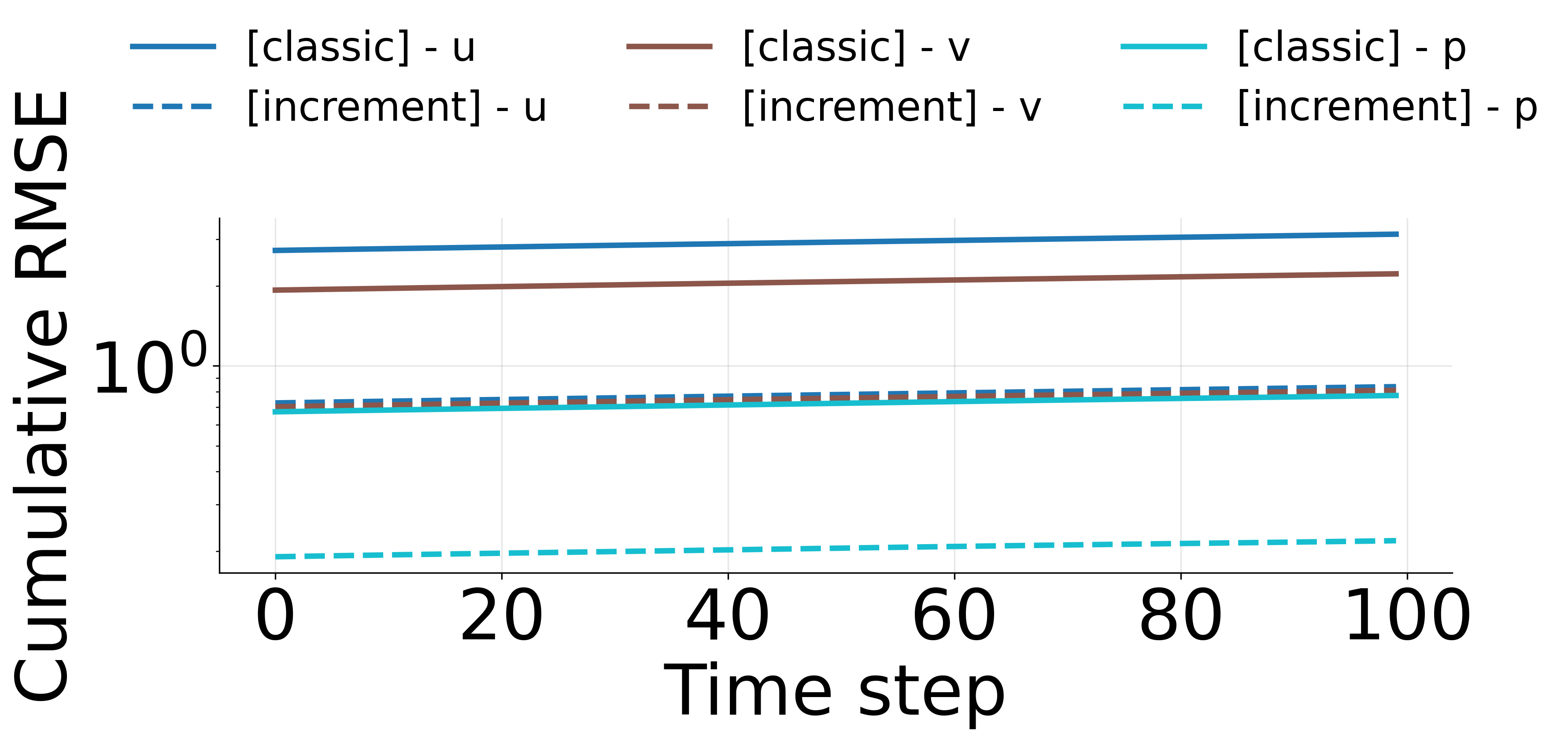}
    \subcaption{}
    \end{minipage}
     \hfill
    \begin{minipage}{0.30\textwidth}
        \centering
        \includegraphics[width=\linewidth]{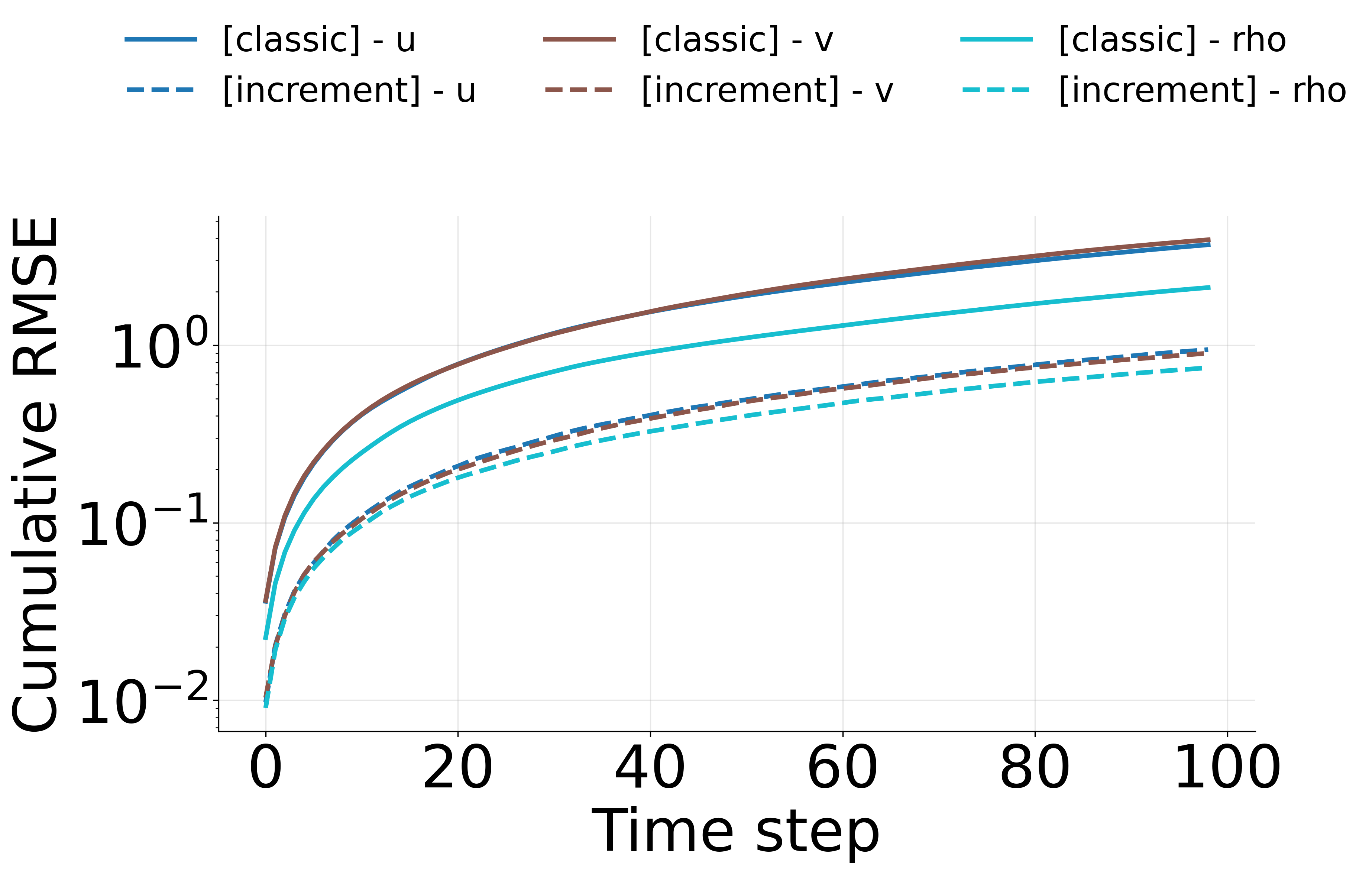}
        \subcaption{}
    \end{minipage}
    \hfill
    \begin{minipage}{0.34\textwidth}
        \centering
        \includegraphics[width=\linewidth]{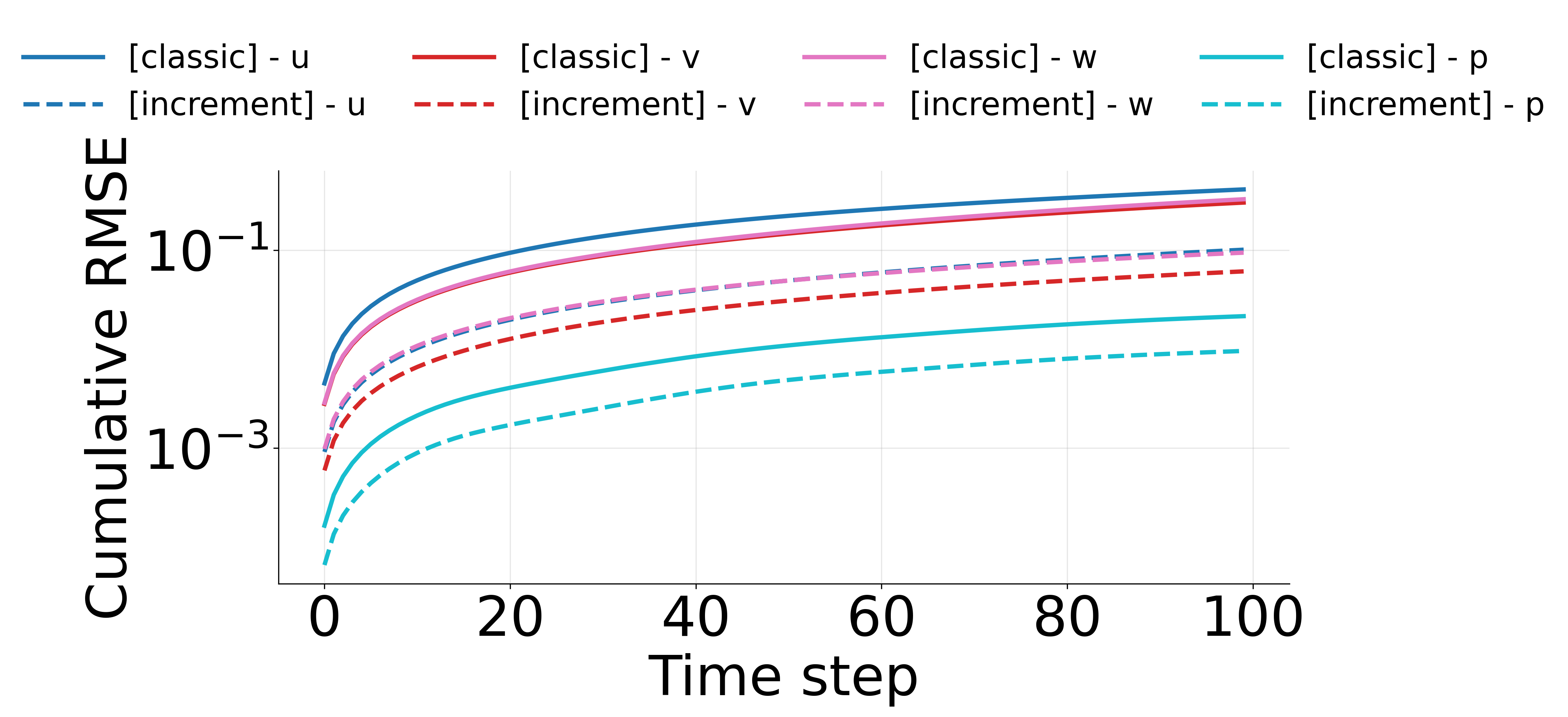}
        \subcaption{}
    \end{minipage}
    \caption{
    Comparison of \textbf{normalized cumulative RMSE} over the last 100 timesteps of the test dataset : 
    Test case 1 by $(u,v)/U_{\mathrm{\infty}}$, Test case 2 by $(u,v)/U_{\mathrm{\infty}},\ \rho/\rho_{\max}$, Test case by 3 $p/ \rho U_{\mathrm{\infty}}^{2}$; the panels (a)-(c) correspond, respectively, to the test case 1 to 3.
    }

    \label{fig:model_accuracy}
\end{figure}

\paragraph{Long-horizon accuracy}
On extended rollouts (\textit{last 100 time steps of the test dataset}), the incremental formulation consistently yields smaller RMSE and a slower rate of error growth across all physical fields, evidencing a marked reduction in autoregressive drift relative to classic targets. By reframing the task as per-step corrections, the method constrains bias accumulation at each horizon step, resulting in more stable long-term predictions and tighter error envelopes, as illustrated in \Cref{fig:model_accuracy}.

We quantify the relative gain of the incremental target as
\begin{equation}
 \Delta(\%) = 100*\,\left[\frac{\mathrm{RMSE}_{\text{inc}} - \mathrm{RMSE}_{\text{classic}}}{\mathrm{RMSE}_{\text{classic}}}\right]  
\end{equation}

Across all test cases (test cases 01–03) and horizons, the incremental target consistently outperforms the classic target: the signed difference \(\Delta\) is negative everywhere, with \textbf{reductions spanning \(54.5\%\)–\(84.2\%\)} as shown in Table~\ref{tab:accum_err}. The benefit is sustained as the horizon grows, for example, test case 01 remains near \(-83\%\) from 10 to 750 steps, and test case 02 strengthens from \(-74.9\%\) at 10 steps to \(-84.2\%\) at 100 steps—indicating markedly lower compounding drift. Even in the more challenging, test case 03 (3D turbulent channel flow), improvements of \(60.5\%-54.5\%\) are maintained out to 100 steps. In absolute terms, the incremental RMSE is typically \textbf{\(2\!\times\)–\(6\!\times\) smaller} than the classic counterpart, reinforcing that learning per-step updates curbs error accumulation and stabilizes long-horizon forecasts.

\begin{table}[h]
\centering
\caption{Comparison of mean accumulated errors after different time horizons for classic vs.\ incremental targets}
\footnotesize
\setlength{\tabcolsep}{3pt}
\renewcommand{\arraystretch}{1.1}
\begin{tabularx}{\linewidth}{@{}l c c c c@{}}
\toprule
\textbf{Case} & \textbf{Horizon (steps)} & \textbf{Classic RMSE} & \textbf{Incremental RMSE} & \(\boldsymbol{\Delta}\) \textbf{(\%)} \\
\midrule
{Test case 01} & 10 &  6.392e-2 &  1.094e-2 &  \textbf{-82.89} $\downarrow$ \\
        & 375 &  2.3961 &  0.4125& \textbf{-82.79} $\downarrow$\\
        & 750 &  4.7961 & 0.8234 & \textbf{-82.83} $\downarrow$\\
\midrule
{Test case 02} & 10 & 0.3372 &  8.463e-2 & \textbf{-74.90} $\downarrow$\\
        & 50 &  2.0446 & 0.45 & \textbf{-77.99} $\downarrow$\\
        & 100 &  6.5170 & 1.0289 & \textbf{-84.21} $\downarrow$\\
\midrule
{Test case 03} & 10 & 0.7819 & 0.3092  &  \textbf{-60.46} $\downarrow$\\
        & 50 & 3.9668 & 1.7497  & \textbf{-55.89} $\downarrow$\\
        & 100 &  7.8828 & 3.5847 & \textbf{-54.53} $\downarrow$\\
\bottomrule
\end{tabularx}
\label{tab:accum_err}
\end{table}

\subsection{Transient dynamics' capture}
\begin{figure}[t!]
    \centering

    \begin{minipage}{0.9\textwidth}
        \subcaptionbox{}[0.1\textwidth]{}
        \hfill
        \begin{minipage}{0.95\textwidth}
            \includegraphics[width=\linewidth]{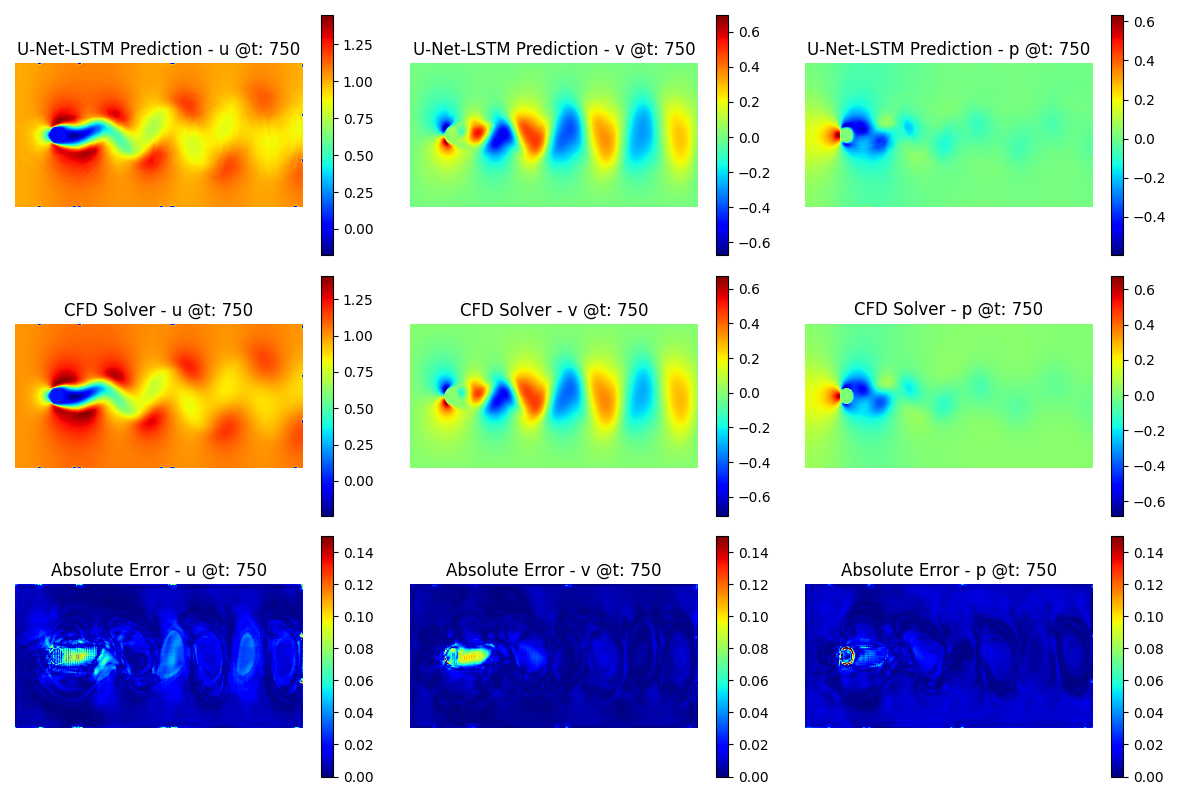}
        \end{minipage}
    \end{minipage}

    \vspace{0.6cm}

    \begin{minipage}{0.9\textwidth}
        \subcaptionbox{}[0.1\textwidth]{}
        \hfill
        \begin{minipage}{0.95\textwidth}
            \includegraphics[width=\linewidth]{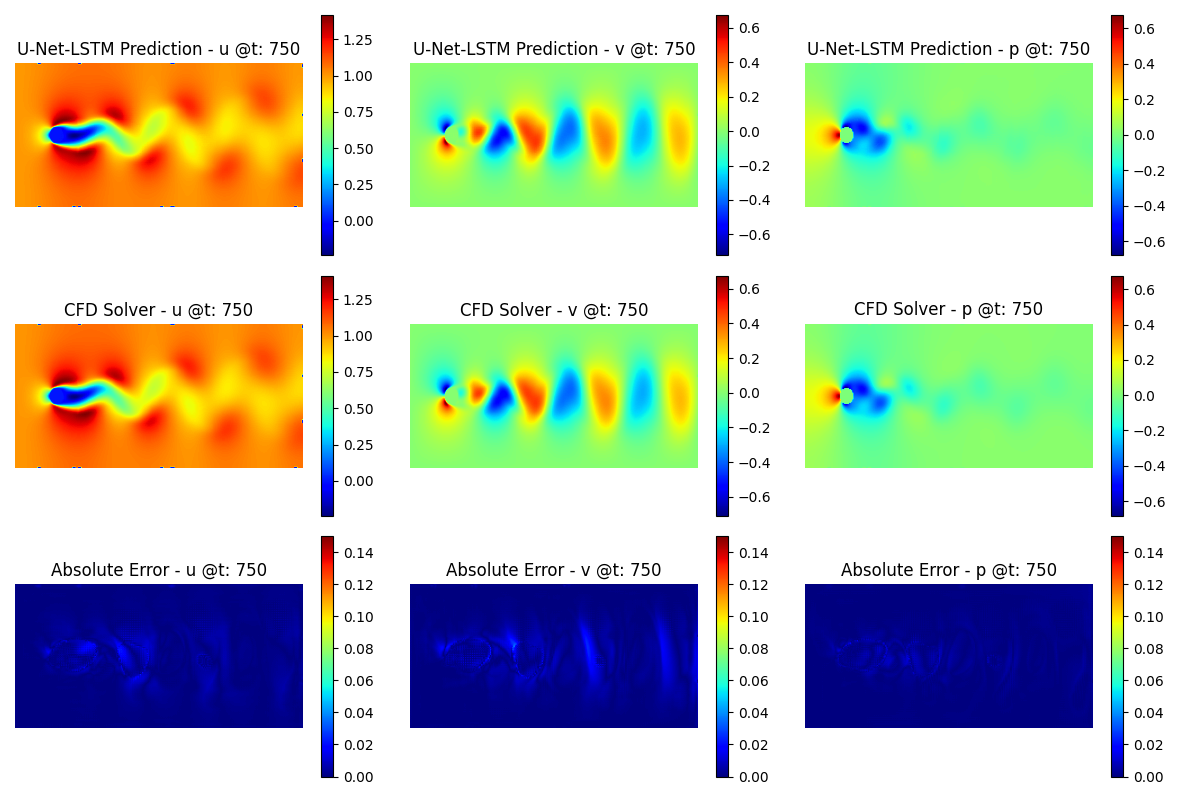}
        \end{minipage}
    \end{minipage}

    \caption{
        Comparison between \textbf{classic} (a) and \textbf{incremental} (b) predictions for Test Case~01 (cylinder wake) at timestep \(t{=}750\).
        Each panel shows the prediction (top row), the CFD reference (middle row), and the absolute-error maps \(|\hat{\phi}-\phi|\) (bottom row) for \(u\), \(v\), and \(p\).
    }
    \label{fig:tc01_compare_vertical}
\end{figure}

\begin{figure}[t!]
    \centering

    \begin{minipage}{0.9\textwidth}
        \subcaptionbox{}[0.1\textwidth]{}
        \hfill
        \begin{minipage}{0.95\textwidth}
            \includegraphics[width=\linewidth]{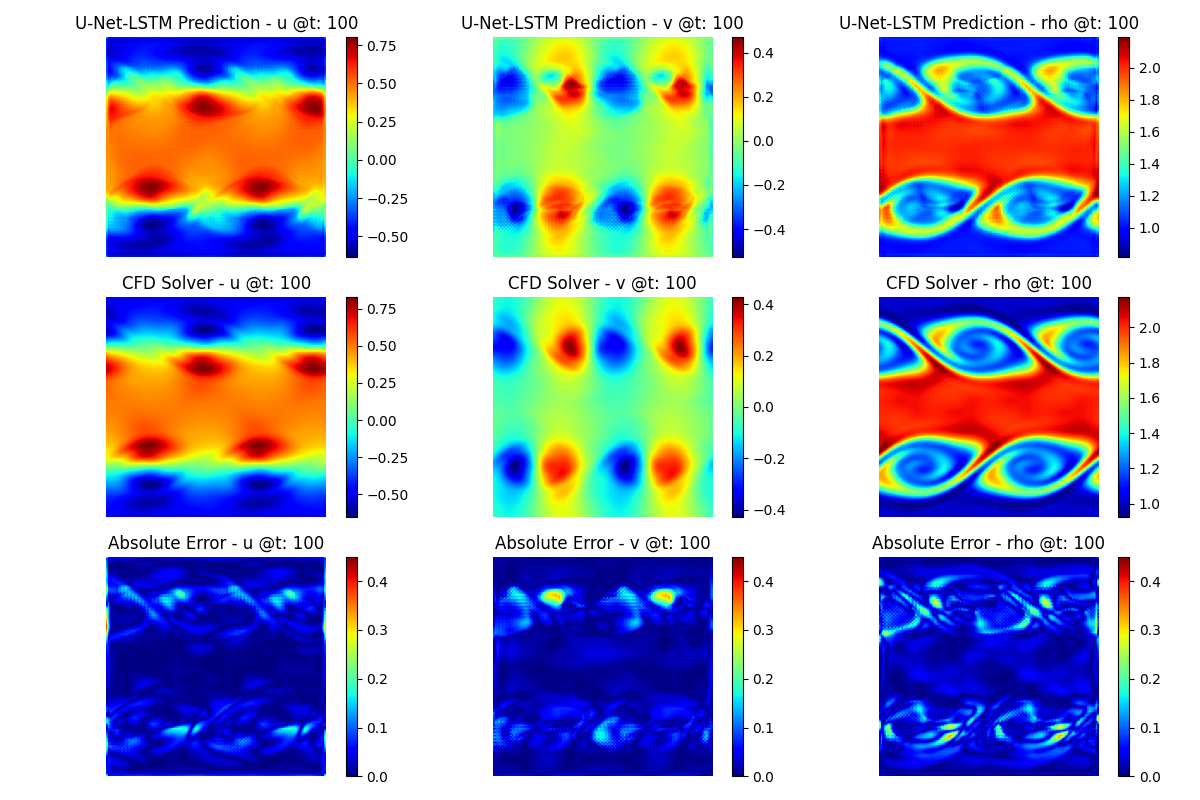}
        \end{minipage}
    \end{minipage}

    \vspace{0.6cm}

    \begin{minipage}{0.9\textwidth}
        \subcaptionbox{}[0.1\textwidth]{}
        \hfill
        \begin{minipage}{0.95\textwidth}
            \includegraphics[width=\linewidth]{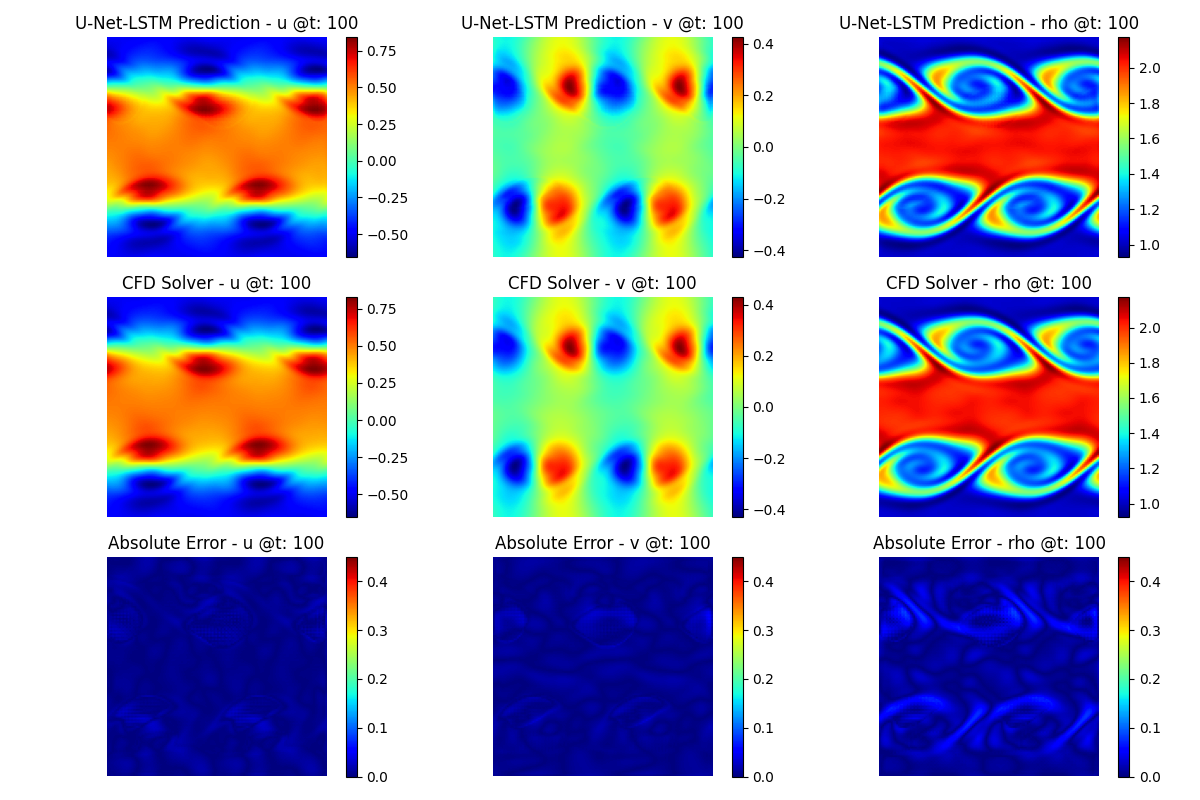}
        \end{minipage}
    \end{minipage}

    \caption{
    Test case 02 (Kelvin–Helmholtz instability) — \textbf{classic} (a) and \textbf{incremental} (b) predictions at \(t{=}100\).
    Visualized fields: \(u\), \(v\), and \(\rho\).  
    Rows follow the same convention as Test Case 01: prediction, CFD reference, and absolute-error maps \(|\hat{\phi}-\phi|\).
    }
    \label{fig:tc02_compare_vertical}
\end{figure}

\begin{figure}[t!]
    \centering

    \begin{minipage}{0.9\textwidth}
        \subcaptionbox{}[0.1\textwidth]{}
        \hfill
        \begin{minipage}{0.85\textwidth}
            \includegraphics[width=\linewidth]{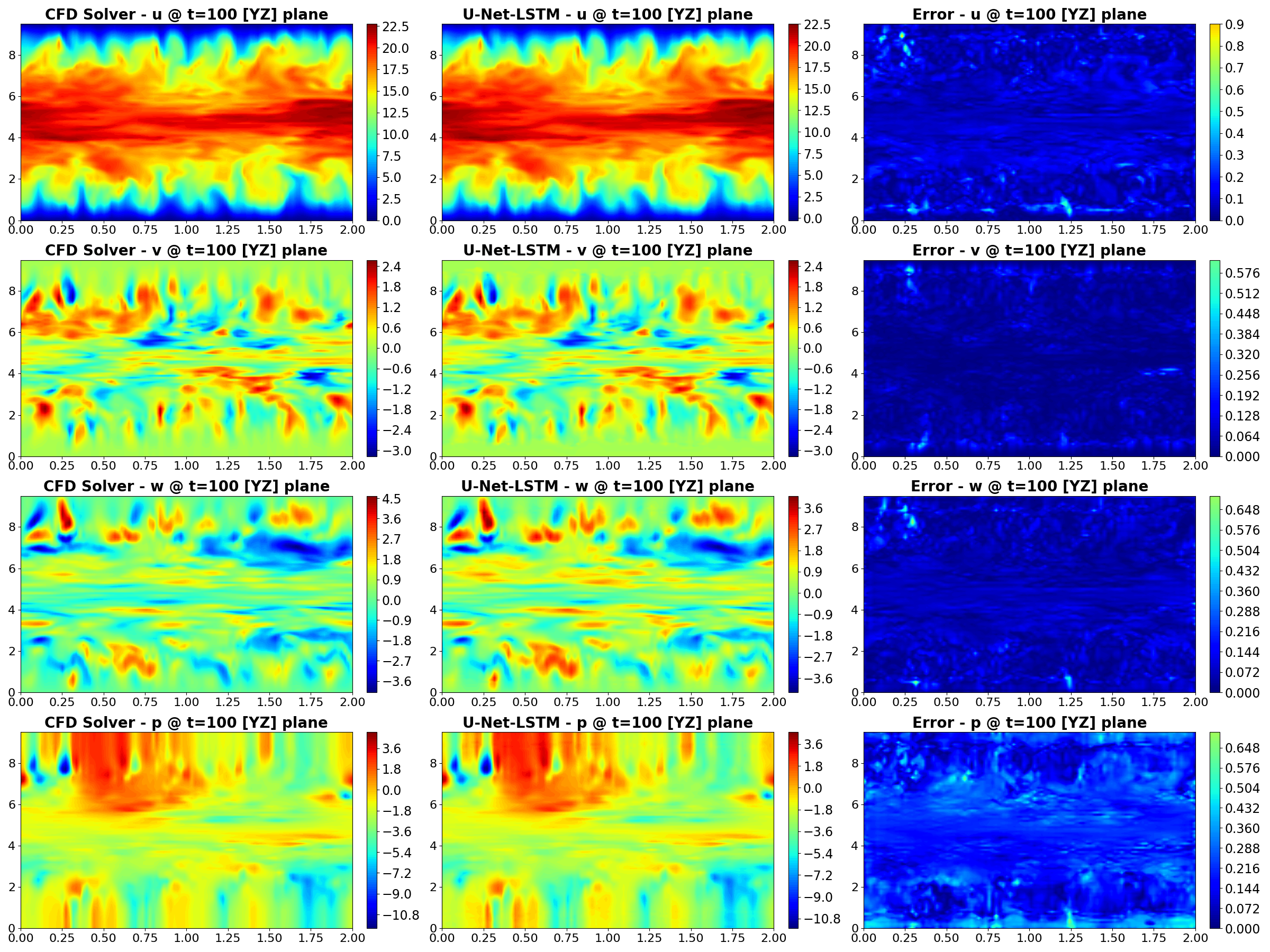}
        \end{minipage}
    \end{minipage}

    \vspace{0.6cm}

    \begin{minipage}{0.9\textwidth}
        \subcaptionbox{}[0.1\textwidth]{}
        \hfill
        \begin{minipage}{0.85\textwidth}
            \includegraphics[width=\linewidth]{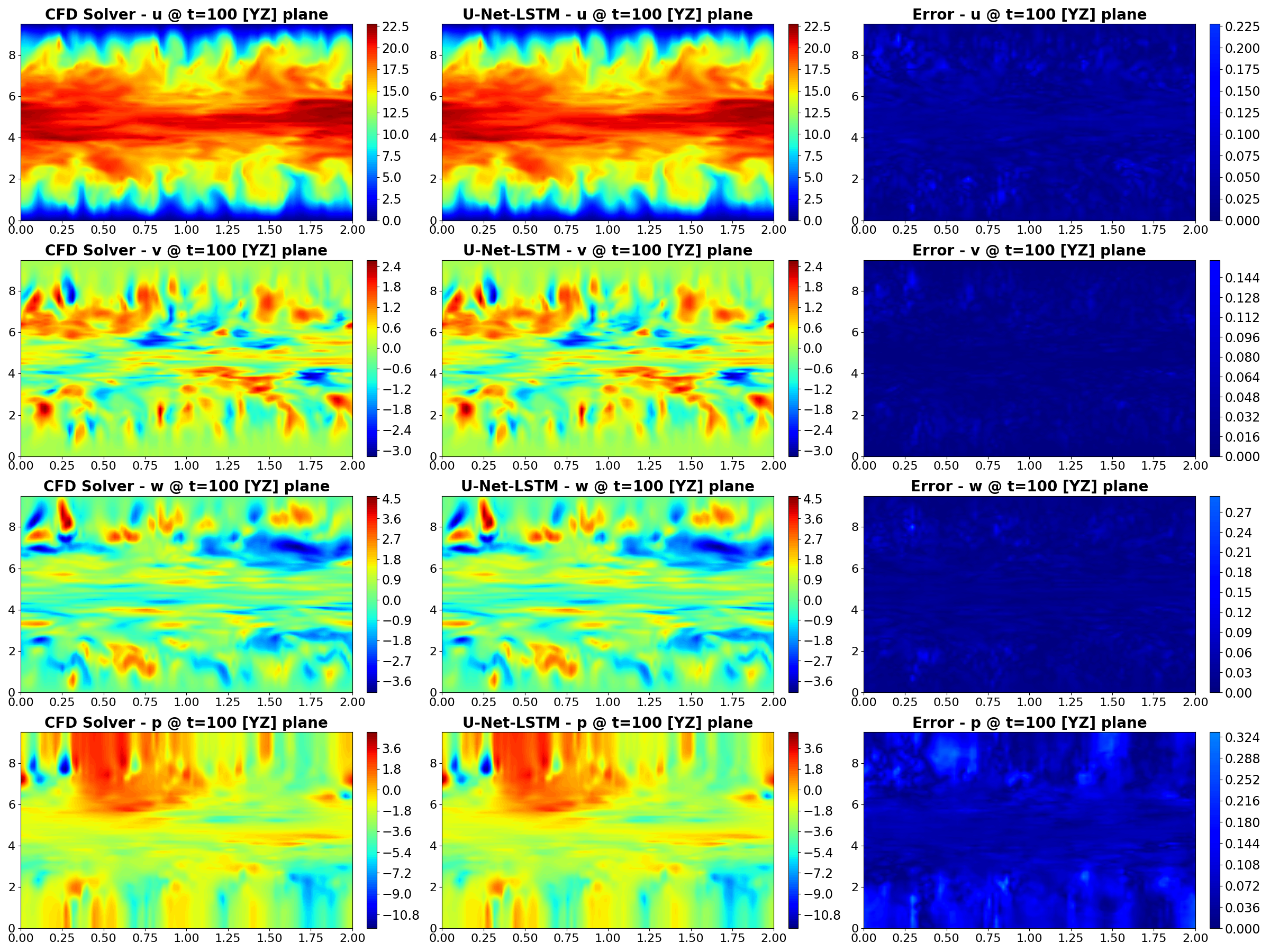}
        \end{minipage}
    \end{minipage}

    \caption{
    Test case 03 (fully developed 3D turbulent channel) — \textbf{classic} (a) and \textbf{incremental} (b) predictions at \(t{=}100\).
    Shown fields: \(u\), \(v\), \(w\), and \(p\).  
    Columns: CFD reference (left), prediction (middle), absolute-error maps (right).
    }
    \label{fig:tc03_compare_vertical}
\end{figure}

Figures from ~\ref{fig:tc01_compare_vertical} to \ref{fig:tc03_compare_vertical} provide a side-by-side comparison of absolute predictions (classic targets) and incremental predictions for the last timestep of each test case. Visually, the incremental approach achieves a sharper recovery of near-wake structures and boundary features.

\newpage

\newpage

\newpage

\begin{figure}[h!] 
    \centering
    \includegraphics[scale=0.4]{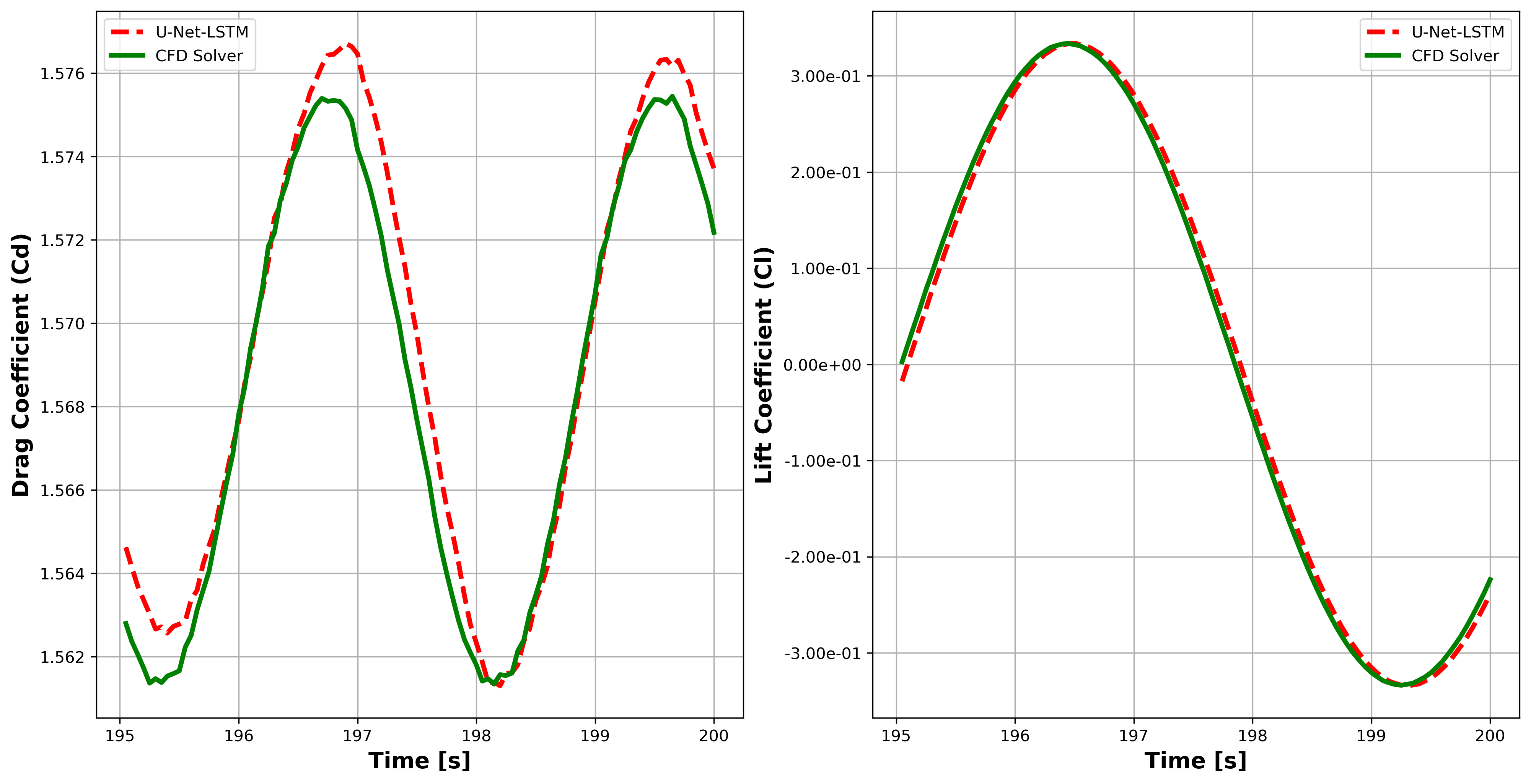}
\caption{Test case 01 (Von Karman Street):
\textbf{Drag and lift coefficients} \((C_d, C_l)\) plot for the last 100 timesteps from the test dataset.
}
\label{fig:tc02_coeffs}
\end{figure}

For the first test case (\textit{cylinder wake}), absolute error maps reveal that the classic model struggles near the cylinder, with larger pressure deviations,  while the incremental model confines errors to narrower regions. Beyond the field-level inspection, the aerodynamic coefficients shown in Figure~\ref{fig:tc02_coeffs} confirm that the incremental formulation preserves the global dynamics of the von Kármán street. The predicted drag and lift coefficients follow the CFD baseline solver with almost identical amplitudes and only a slight phase offset, indicating that the model captures the vortex-shedding cycle and its integral signatures with high fidelity.

In the case of \textit{Kelvin-Helmholtz shear layer}, increments provide a better representation of the roll-up and merging of vortical structures, maintaining closer agreement in both velocity and density fields. 

For the more complex, the test case 03 (\textit{turbulent 3D channel}), although both models show difficulty in fully reconstructing fine-scale turbulence, the incremental predictions exhibit reduced error intensity across all components (\(u,v,w,p\)), yielding a visibly closer match to CFD reference fields. Overall, the field-level inspection confirms that incremental learning mitigates localized drift, preserves coherent structures, and provides more physically consistent solutions across diverse flow regimes.

\begin{figure}[h!] 
    \centering
    \includegraphics[scale=0.45]{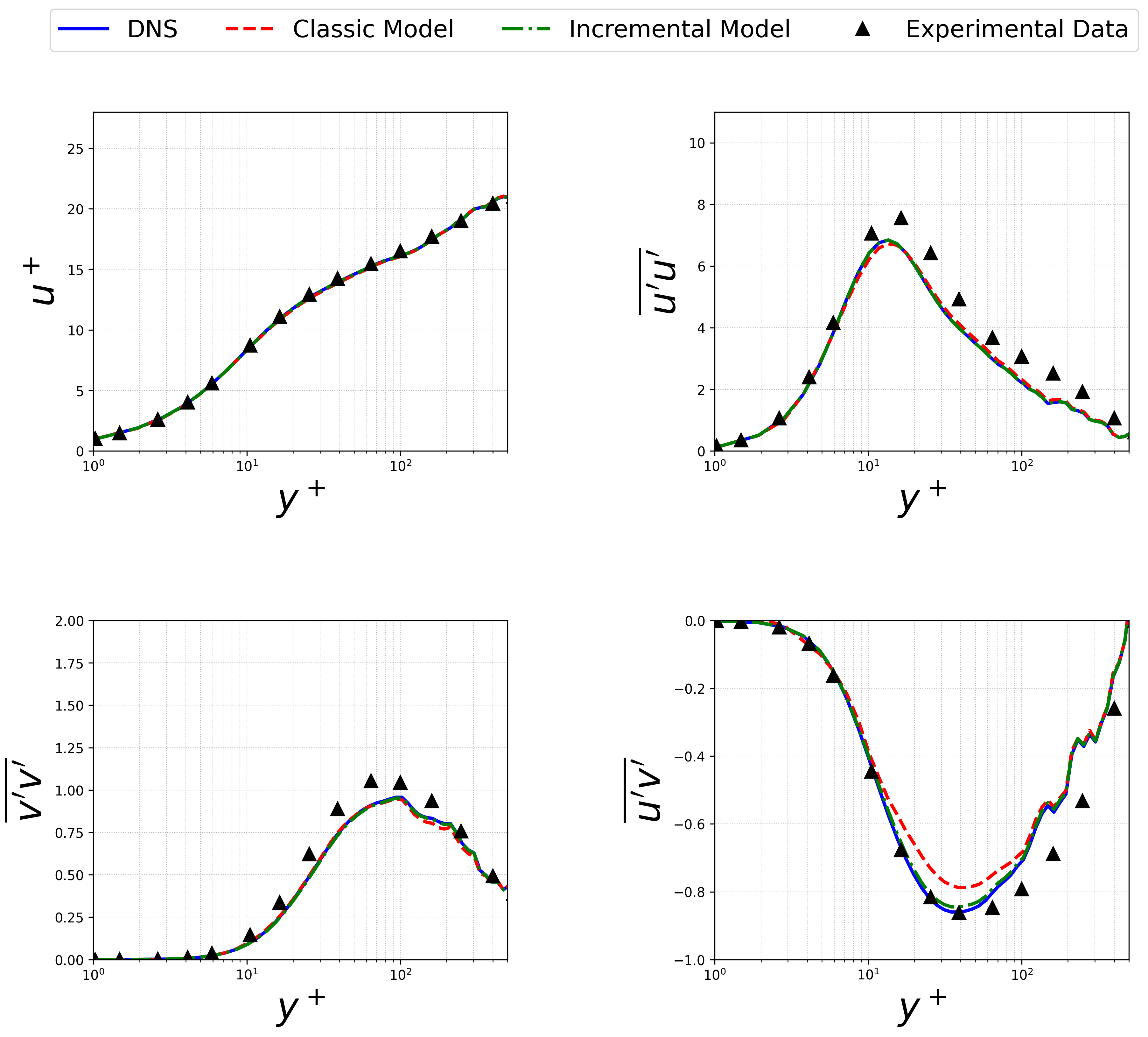}
\caption{Test case 03 (turbulent channel): wall-normal statistics in wall units \(y^+\).
(a) mean velocity \(u^+(y^+)\);
(b) streamwise normal stress \(\overline{u'u'}\);
(c) wall-normal normal stress \(\overline{v'v'}\);
(d) Reynolds shear stress \(\overline{u'v'}\).
}
\label{fig:tc03_reynolds_stresses}
\end{figure}

This qualitative advantage is corroborated quantitatively by the wall-normal statistics in Figure~\ref{fig:tc03_reynolds_stresses}: the incremental model follows DNS and experiment more closely across the buffer and logarithmic regions (e.g., \(10 \lesssim y^+ \lesssim 50\) and \(y^+ \gtrsim 50\)), reduces the near-wall peak bias in \(\overline{u'u'}\) and \(\overline{v'v'}\), and improves both the magnitude and location of the \(\overline{u'v'}\) minimum, while also tightening agreement for the mean profile \(u^+(y^+)\).

Taken together, the field maps and wall-normal statistics indicate that incremental learning mitigates localized drift, preserves turbulent structures, and better captures momentum transfer mechanisms across regimes.

\begin{figure}[h!] 
    \centering
    \includegraphics[scale=0.38]{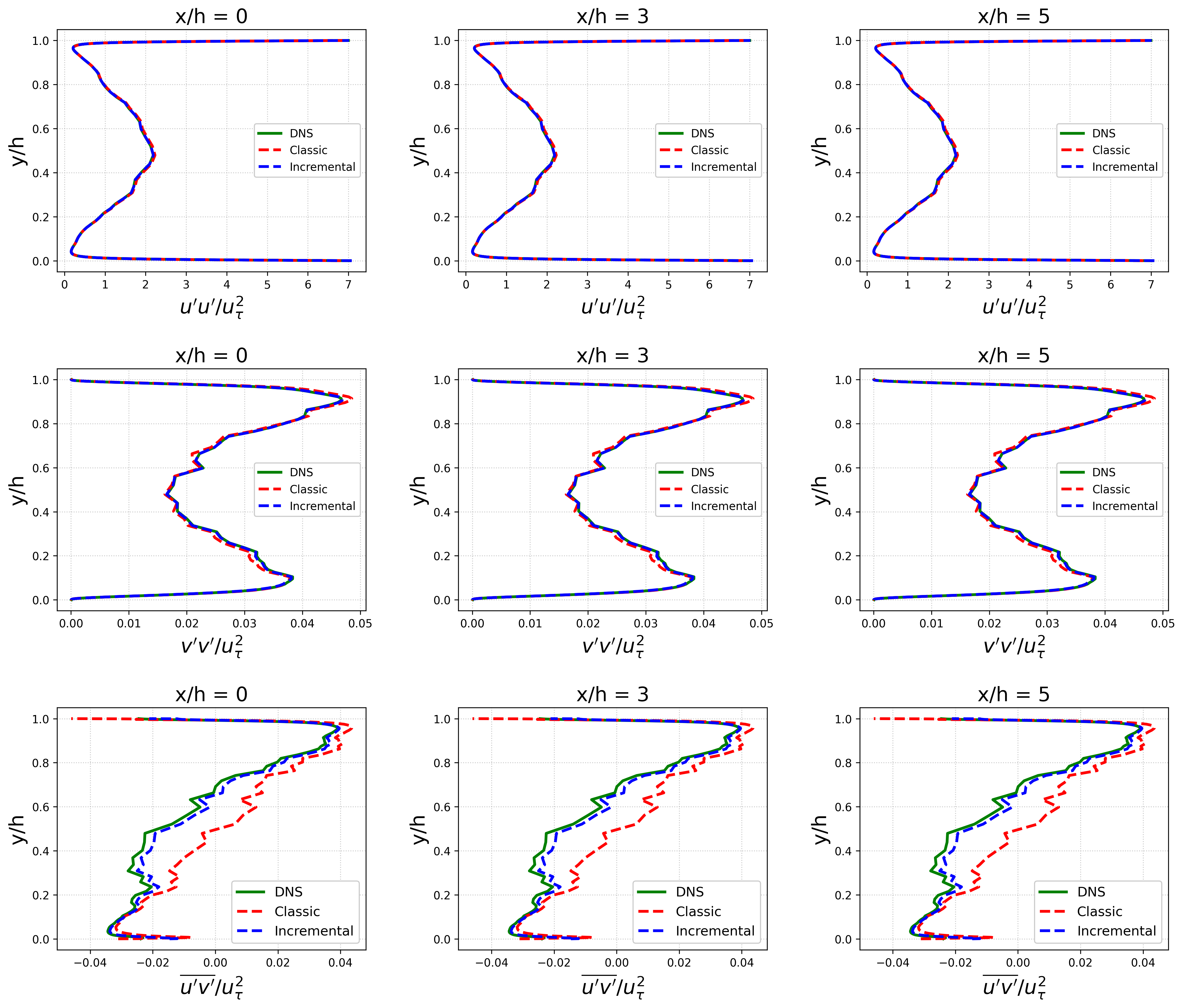}
\caption{Test case 03 (turbulent channel): spanwise-averaged Reynolds-stress profiles vs.\ wall-normal coordinate \(y/h\) at streamwise stations \(x/h=0,3,5\).
Rows: (top) \(\overline{u'u'}/u_\tau^{2}\), (middle) \(\overline{v'v'}/u_\tau^{2}\), (bottom) \(\overline{u'v'}/u_\tau^{2}\).
Columns: \(x/h=0\) (left), \(3\) (center), \(5\) (right).
}
\label{fig:tc03_profiles_xh}

\end{figure}

Across all stations in Figure~\ref{fig:tc03_profiles_xh} (\(x/h=0,3,5\)), the incremental formulation tracks DNS more tightly: the near-wall peaks of \(\overline{u'u'}/u_\tau^2\) and \(\overline{v'v'}/u_\tau^2\) align in both height and wall-normal location, and the \(\overline{u'v'}/u_\tau^2\) minimum appears at the correct \(y/h\) with improved magnitude.

By contrast, the classic model tends to overpredict normal-stress peaks and displace the shear-stress minimum outward, with greater station-to-station variability.

\newpage
\subsection{Memory resources management using incremental learning}

\begin{figure}[h!] 
    \centering
    \includegraphics[scale=0.3]{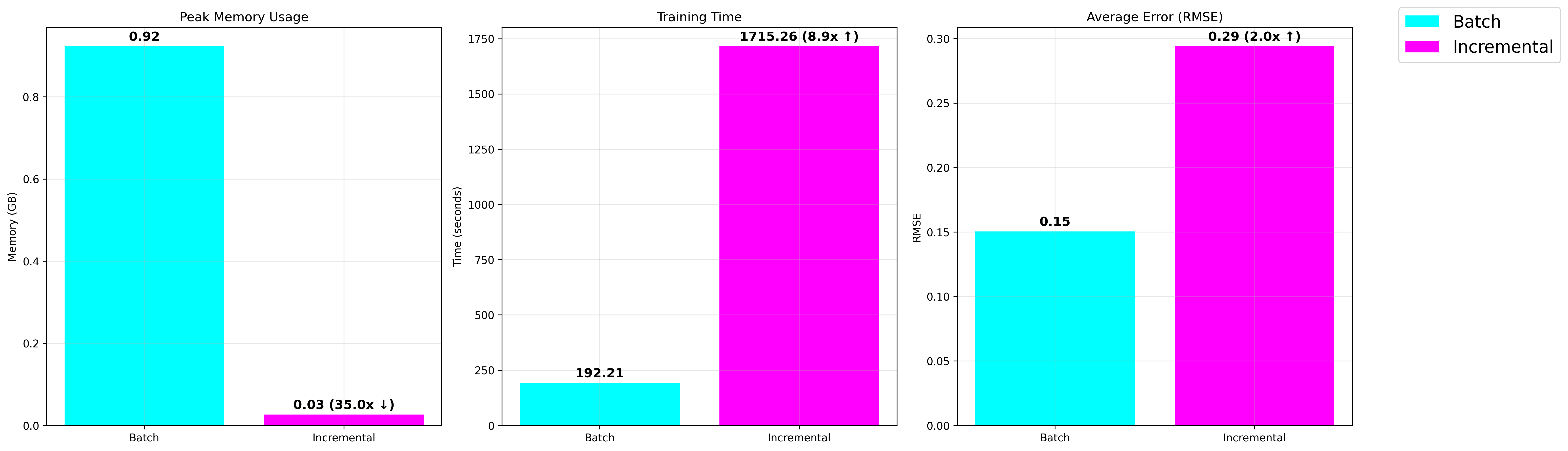}
\caption{Comparison batch vs incremental training metrics
}
\label{fig:tc03_memory}
\end{figure}

The resource-usage comparison in Figure ~\ref{fig:tc03_memory} highlights a clear but non-trivial trade-off introduced by incremental learning. On the one hand, the incremental strategy reduces peak GPU memory from $  0.92~\text{GB} $ to $  0.03~\text{GB} $, a factor of $ 35\times $, which is critical for 3D DNS/LES data, where storing many snapshots simultaneously is prohibitive. This makes it possible to train on long time sequences and high-resolution fields that would simply not fit in memory under a conventional batch regime. On the other hand, this memory relief is obtained at the cost of an $  8.9\times $ increase in wall-clock training time (from $ 192~\text{s} $ to $  1715~\text{s} $) and a moderate degradation in accuracy (RMSE rising from $0.15$ to $0.29$, i.e.\ about $2\times$).

From a critical perspective, these results show that incremental learning is not a universally superior option but rather a pragmatic compromise: it trades part of the training efficiency and predictive sharpness for a substantial gain in scalability. The approach becomes particularly relevant when training represents a one-time offline cost that is later amortized over many CFD–ML rollouts or when memory availability is the primary limitation (e.g., large 3D channels, extensive parametric studies, in-situ training on constrained GPUs). In settings where computational resources are ample and rapid experimentation or hyperparameter optimization is the main objective, classic batch training remains the more effective strategy. Incremental learning should, therefore, be regarded as a resource-management tool that expands the range of feasible problem sizes rather than as a direct performance enhancer. Its adoption ultimately depends on the hardware budget and the level of predictive accuracy required by the downstream hybrid solver.

\section{Conclusion}
\label{sec:conclusion}

This work addressed a central limitation of data-driven surrogates for unsteady CFD: most models either drift over long rollouts or require prohibitive memory and data to remain accurate. Our objective was to design a spatiotemporal predictor that (i) remains stable over long horizons, (ii) can be trained on realistic DNS/LES datasets with limited hardware, and (iii) fits naturally into segregated finite–volume solvers such as PISO and PIMPLE for transient simulations.

To this end, we introduced an Incremental Time-Stepping U-Net–LSTM that predicts field updates instead of absolute states. By aligning learning with classical time-marching schemes, the model reduces autoregressive drift and better preserves the underlying dynamics. Across three representative test cases (2D cylinder wake, Kelvin–Helmholtz shear layer, and 3D turbulent channel), the incremental targets consistently lower long-horizon RMSE relative to classic absolute-field supervision, with reductions in accumulated error ranging from \(\approx 54.5\%\) to \(84.2\%\) (Table~\ref{tab:accum_err}). Field-level comparisons and turbulence statistics further show that the incremental formulation captures coherent structures and Reynolds-stress profiles more faithfully than the baseline.

The study also examined resource usage in a realistic training setting. Incremental learning drastically reduces peak GPU memory (by a factor of \(35\times\)), enabling training on long sequences and volumetric fields that would be infeasible with conventional batch loading (Figure ~\ref{fig:tc03_memory}). This gain comes at the cost of slower training (about \(8.9\times\) larger wall-clock time) and a moderate increase in RMSE. These results highlight that incremental learning is not a free improvement but a resource-management strategy: it trades training speed and a small amount of accuracy for scalability and the ability to handle large-scale DNS/LES databases on commodity hardware.

Overall, the main contributions of this work can be summarized as follows:
\begin{itemize}
    \item An incremental U-Net–LSTM formulation is proposed, predicting per-step increments to mitigate long-horizon error accumulation and enhance temporal stability during extended rollouts.
    
    \item Across three canonical test cases, the incremental formulation achieves a reduction of \(54.5\%\)–\(84.2\%\) in cumulative prediction errors relative to classic absolute-field targets, while maintaining essential engineering metrics such as drag/lift coefficients and Reynolds-stress profiles.
    
    \item An incremental training strategy is introduced to significantly reduce memory requirements, thereby clarifying the trade-off between computational cost, prediction accuracy, and scalability for large-scale unsteady flow datasets.
    
    \item The resulting model is explicitly tailored for integration into segregated CFD solvers, where it can serve either as a refined initial guess or as a corrective module for pressure–velocity coupling within PISO/PIMPLE algorithms.
\end{itemize}

Future work will focus on deploying the proposed model in a fully coupled hybrid CFD–ML framework. In particular, we will embed the Incremental Time-Stepping U-Net–LSTM into PISO/PIMPLE loops as a warm-start predictor for pressure and velocity fields, and as a corrector that refines provisional states before the pressure projection. This setting will allow us to measure net wall-clock gains, including both inference and solver overheads, and to assess robustness during long transient runs. Additional directions include coupling the network to residual-based loss terms from the CFD solver, exploring adaptive time stepping informed by the learned increments, and testing generalization across Reynolds numbers, geometries, and boundary conditions. These steps are essential for turning the present surrogate into a practical building block for efficient hybrid CFD-ML solvers for industrial-scale transient flows.


\end{document}